%% file: notex_v2.tex
\documentclass[twocolumn,english,pra]{revtex4-1}
\usepackage{lmodern}
\usepackage[T1]{fontenc}
\usepackage[latin9]{inputenc}
\usepackage{amsmath}
\usepackage{amssymb}
\usepackage{graphicx}
\usepackage{babel}
\begin{document}
\title{$p$-band stability of ultracold atom gas in anharmonic optical lattice
potential with large energy scales}
\author{Mateusz \L \k{a}cki}
\affiliation{Institute of Theoretical Physics, Jagiellonian University, \L ojasiewicza
11, 30-348 Kraków, Poland}
\begin{abstract}
Using an optical potential with subwavelength resolution in the form
of sharp $\delta$-like peaks, potential landscapes are created with
increased anharmonicity in placement of lattice band energies and
more favorable energy scales. In particular, this makes the ultracold
atom $p$-band gas more stable. The article outlines the details of
the construction and discusses the $p$-band stability in canonical
cosine optical lattice potential, double well potential, and a combination
of a classical cosine potential with dark state peaked potential.
\end{abstract}
\maketitle

\section{Introduction}

Optical lattices make a convenient and powerful setting for experimental
study of many-body physics using ultracold atoms. An observation that
such systems are described by Hubbard-type models \citep{Jaksch1998},
followed by a realization of the Mott insulator-superfluid quantum
phase transition \citep{Greiner2002}, led to numerous proposals and
experiments with ultracold atoms \citep{Lewenstein2012}. 

One of the research directions was populating higher bands of optical
lattices \citep{Muller2007,Wirth2011} with the ongoing debate on
stability of such systems. This was in part motivated by prospects
to create interesting superfluid states, time reversal symmetry breaking
\citep{Li2012,Sowinski2013} or emergence of $p_{x}+ip_{y}$ order
\citep{Hauke2011,Olschlager2013}, or quantum Hall physics \citep{Wu2008,Zhang2011}.

In addition to theory proposals, questions of more practical significance
have been raised -- a problem of preparing gas in the excited bands
\citep{Gemelke2005,Sowinski2012,acki2013,Zhou2018} and the question
of stability and lifetime of such a gas \citep{Kohl2005,Muller2007,Hu2015,Niu2018}. 

Higher bands can also be populated by coherent resonant band coupling
\citep{Gemelke2005,Sowinski2012,acki2013,Cabrera-Gutierrez2019}.
When the coupling of $s$ to $p$ band is resonant, and presence of
other bands can be disregarded, a synthetic-dimension two-leg ladder
system, carrying flux $\pi$ per ladder plaquette is created \citep{Strater2015}.

In the standard optical lattice, the collisional stability of the
$p$-band e.g. is limited due to the fact that total energy of two
particles is $p$ band is often very close to the configuration where
one particle is in $s$ and the other in the $d$ band. This process
is off-resonant when the lattice band energies are anharmonic, leading
to prolonged lifetime, as observed in the experiment \citep{Kastberg1995,Isacsson2005,Muller2007}.

Deviation from equal spacing between subsequent band $s,p,d,\ldots$,
necessary for collisional $p$-band stability has to dominate other
energy scales such as interaction strength or amplitudes of time-dependent
fields. The latter can be lowered, thus increasing the stability.
The price to pay however would be limits on simulable physics, given
practical limits on coherence time of ultracold lattice systems. The
real solution should be in the direction of making energy levels in
lattice systems more ahnarmonic, and the energy scales larger. 

In \citep{Lacki2016,Wang2018} a method of creating a potential in
the form of a comb of subwavelength, few-nanometer wide peaks was
proposed. It carries bands with non-harmonic energies $\sim n^{2}$.
Moreover subwavelength-width double well systems \citep{Budich2017}
allow large values of energy scales for hopping and interaction.

In this work we explore the effects the anharmonic level spacing has
on collisional stability of $p$-band gas and on coherent resonant
coupling of $s$ and $p$ bands. We also study the energy scales for
the parameters of tight-binding models describing motion of ultracold
atom in such potentials.

In Section II we review the tight-binding description of an ultracold
atom gas in few lowest Bloch bands of the optical potential. We discuss
energy level arrangement for various particular potentials: ``standard''
$\cos^{2}$ optical lattice, the double well lattice, subwavelegth
comb potential, and a combination of comb potential and the standard
$\cos^{2}$ lattice. In Section III we discuss the simulation of long-term
depletion of the $p$-band by collisional interactions in case for
all considered optical potentials. In Section IV, we study the creation
of the synthetic-dimension two-leg ladder system in $s$ and $p$
band of a 1D lattice system, focusing on the achieved energy scales
and the containment of the system. We provide summary and outlook
in Section V.

\section{Multiband description of the ultracold atom gas in the optical lattices}

\label{sec:potentials}

The gas of ultracold atoms of mass $m_{a}$ in the periodic optical
potential $V_{\textrm{opt.}}(r)$ is canonically described by the
second quantization Hamiltonian of the form \citep{Jaksch1998}:

\begin{eqnarray}
H_{X} & = & \int\textrm{d}^{3}r\hat{\psi}^{\dagger}(r)\left[-\frac{\hbar^{2}}{2m_{a}}\nabla^{2}+V_{\textrm{opt.}}(r)\right]\hat{\psi}(r)\nonumber \\
 &  & +\frac{g}{2}\int\hat{\psi}^{\dagger}(r)\hat{\psi}^{\dagger}(r)\hat{\psi}(r)\hat{\psi}(r)\textrm{d}^{3}r.\label{eq:2ndquantizH}
\end{eqnarray}
The $g=\frac{4\pi a_{s}\hbar^{2}}{m_{a}}$ is the strength of two-particle
collisional interactions by $s$-wave scattering with a scattering
length $a_{s}$, tunable by Feshbach resonances \citep{Chin2010}.
Tight harmonic confinement $m_{a}\Omega^{2}(y^{2}+z^{2})/2$ in $y$
and $z$ makes the system effectively 1D. 

The rest of this Section is organized as follows. In the Subsection
\ref{subsec:Tight-binding-description} we restate the multiband tight-binding
description of (\ref{eq:2ndquantizH}). In Subsection \ref{subsec:Optical-potentials}
we briefly introduce the four potentials that will be compared for
the $p$-band stability using the quantity $f$ which is introduced
in the second part of this Subsection. In Subsections \ref{subsec:classicalOpticalLattice}-\ref{subsec:Classical-potential-together}
we discuss the particular quantitative features of the four potentials
considered in this work.

\subsection{Tight-binding description}

\label{subsec:Tight-binding-description}

In this section we recapitulate the conventional multi-band tight-binding
decription of the Hamiltonian (\ref{eq:2ndquantizH}).

The $x-$periodic potential, $V_{\textrm{opt.}}(\vec{r})=V_{\textrm{opt.}}(x)+m_{a}\Omega^{2}(y^{2}+z^{2})/2$
admits a family of $x$-quasiperiodic eigenfunctions $B_{k}^{\alpha}(\vec{r})=B_{k}^{\alpha}(x)\mathcal{H}(y)\mathcal{H}(z),B_{k}^{\alpha}(x+a)=B_{k}^{\alpha}(x)e^{ika/\hbar}$
for each band $\alpha=0,1,2,\ldots$ (with $\alpha=1$ corresponding
to the $p$ band) to the eigenenergy $E^{\alpha}(k)$. The $k$ is
a quasimomentum taken from the Brillouin zone $k\in[-\pi/a,\pi/a]$,
$a$ is the lattice constant, and $\mathcal{H}(\cdot)$ is the harmonic
potential ground state. For large enough $\Omega$ just the lowest
mode in $y,z$ is populated which reduces down to global energy shift.
In this work we consider only $V_{\textrm{opt.}}(x)$ which are smooth
and have one or two minima in one period.

For isolated bands, for each lattice site $x_{n}$, exponentially
localized Wannier functions $W_{n}^{\alpha}(x)$ can be constructed
under mild assumptions \citep{Kohn1959,Kivelson1982,Marzari2012}.
Specifically:

\[
W_{n}^{\alpha}(x)=N\int_{k\in BZ}B_{k}^{\alpha}(x)e^{i(\theta_{k}+kan)}\textrm{d}k,
\]
where $\theta_{k}$ is chosen to minimize the spatial variance of
the Wannier functions, and $N$ ensures unit $L^{2}$ norm of $W_{n}^{\alpha}$'s
\citep{Kohn1959,Kivelson1982,Marzari2012}.

When the Hamiltonian (\ref{eq:2ndquantizH}) is expressed in the basis
$W_{k}^{\alpha}(x)$, the Multi-band Bose-Hubbard (MBH) Hamiltonian
is obtained

\begin{eqnarray}
H_{\textrm{MBH}} & = & -\sum\limits _{\alpha,n,m}J_{nm}^{\alpha\alpha}(\hat{a}_{n}^{\alpha})^{\dagger}\hat{a}_{m}^{\alpha}+H.c.\nonumber \\
 &  & +\frac{1}{2}\sum\limits _{\stackrel{\alpha\ldots\delta}{n\ldots p}}U_{nmop}^{\alpha\beta\gamma\delta}(\hat{a}_{n}^{\alpha})^{\dagger}(\hat{a}_{m}^{\beta})^{\dagger}\hat{a}_{o}^{\gamma}\hat{a}_{p}^{\delta},\label{eq:MBH}
\end{eqnarray}
where: 
\begin{eqnarray}
J_{nm}^{\alpha\alpha} & = & -\int dx\ W_{n}^{\alpha}(x)\!\!\left(\!-\frac{\hbar^{2}}{2m}\nabla^{2}\!+\!V_{\textrm{opt.}}(x)\!\!\right)\!\!W_{m}^{\alpha}(x)\\
 & =\frac{1}{\textrm{vol}(BZ)} & \int_{BZ}e^{i(n-m)ka}E^{\alpha}(k)\textrm{d}k,\label{eq:Jform}
\end{eqnarray}
and
\begin{equation}
U_{nmop}^{\alpha\beta\gamma\delta}=g\int d\vec{r}\ W_{n}^{\alpha}(\vec{r})W_{m}^{{\beta}}(\vec{r})W_{o}^{{\gamma}}(\vec{r})W_{p}^{{\delta}}(\vec{r}).\label{eq:Uint}
\end{equation}
In particular, $J_{nn}^{\alpha\alpha}\equiv\bar{E}_{n}^{\alpha}$
is the on-site energy of a particle at site $n$ and band $\alpha.$
The $J_{n,n+1}^{\alpha\alpha}\equiv J^{\alpha}$ is nearest-neighbor
hopping rate within band $\alpha$. In the formula for integrals $U_{nmop}^{\alpha\beta\gamma\delta}$,
an integral of $H^{4}(y)H^{4}(z)$ over $y,z$ directions gives an
overall factor that is incorporated in $g$.

When the gas populates the $p$-band, the full model (\ref{eq:MBH})
reduces to a single band Bose-Hubbard model,

\begin{equation}
H_{BH}=-J\sum\limits _{\langle n,m\rangle}(\hat{a}_{n}^{\dagger}\hat{a}_{m}+H.c.)+\sum_{m}\frac{U}{2}\hat{n}_{m}(\hat{n}_{m}-1),\label{eq:BH}
\end{equation}
where $U=U_{nnnn}^{1111},J=J_{n,n+1}^{11}=J^{1}.$ The model exhibits
superfluid-Mott insulator quantum phase transition which occurs for
$J/U\approx0.3$ \citep{Fisher1989,Greiner2002,Stoeferle2004,Zakrzewski2008,Li2011,Mark2011}.
The question of whether the multi-band model $H_{\textrm{MBH}}$ can
be truncated to single band $H_{BH}$ is the central question of this
paper.

The Hamiltonian $H_{\textrm{MBH}}$ describes a complex many-body
system which is non-integrable. It conserves the total number of particles.
Conventionally the truncation of the total Hilbert space is by restricting
the $H_{\textrm{MBH}}$ to lowest few bands. The Hilbert space for
$H_{\textrm{MBH}}$ for gas of $N$ particles occupying $L$ site
lattice within lowest $\alpha_{\textrm{max}}$ bands has dimension
${N+\alpha_{\textrm{max}}L-1 \choose N}$ which is a prohibitively
large number of parameters for representing the Hamiltonian or even
eigenvectors for large systems.

Efficient numerical simulation is not possible in lattices in dimension
higher than 1. The dynamics of the single-band variant, the Bose-Hubbard
model can be effectively simulated in the 1D lattices using the TEBD
or various tensor network approaches. Time-complexity of these methods
is proportional to the $\sim d^{3}$ where $d$ is the single-site
Hilbert space dimension. For $\alpha_{\textrm{max}}=3$, and truncation
at 6 particles per lattice site, we have $d=10$, factor $O(10)$
larger than ``typical'' cutoffs for BH Hamiltonian calculations.
Ultimately the limit of application of these methods for simulation
of the dynamics is growth of entanglement entropies.

In this work, to minimize the computational effort, we restrict our
study to small systems of density $N/L>1$, and to minimize the boundary
effects, perdiodic boundary conditions are used. For $L=6$ and $\alpha_{\textrm{max}}=3$
the full representation of vectors can be used.

\subsection{Optical potentials and $p$-band collisional stability}

\label{subsec:Optical-potentials}

This section provides a short overview of different optical potentials
considered in this work. In the second part we introduce quantitiy
$f$ which is useful for quantifying $p$-band stability of these
potentials.

For ultracold atoms arbitrary optical potentials can can created using
either stroboscopic projection schemes (as proposed in \citep{Lacki2019})
or spatial light modulators (as exemplified in \citep{Boyer2006,Liang2010,Bowman2015}).
Such techniques, however, imply either short system lifetime or relatively
long lattice constants $a\gg\lambda/2$ and in turn overall low energy
scale for lattice dynamics given by the recoil energy $E_{R}=\frac{h^{2}}{8ma^{2}}$.
The reliable workhorse for ultracold lattice systems have been potentials
due to the AC-Stark shift from a laser standing-wave, far detuned
from an optical transition \citep{Jaksch1998}: 
\begin{equation}
V_{\textrm{latt}}(x)=V_{x}\cos^{2}kx,\ \ k=2\pi/\lambda\label{eq:Vlatt}
\end{equation}
where $\lambda$ is the laser wavelength. By using two laser wavelengths
the following potential is obtained \citep{Sebby-Strabley2006}:
\begin{eqnarray}
V_{dw}(x) & \!=\! & V_{2}\cos^{2}(k_{2}x)\!+\!V_{1}\cos^{2}(k_{1}x+\phi),\ k_{1}=\frac{k_{2}}{2}.\label{eq:Vdw}
\end{eqnarray}
The above potential has been promimently used in $V_{2}>V_{1}/4$
regime, where it features two minima per period. It is of main interest
for us for $V_{2}=V_{1}/4$ where it is a lattice potential with unique
minimum per period, which is quartic.

Recently a scheme, based on coherent dark-state of atomic $\Lambda$
system has been proposed \citep{Lacki2016,Wang2018} that allows to
create comb potentials of the form:\\

\begin{equation}
V_{na}(\epsilon,x)=E_{R}\frac{\epsilon^{2}\cos^{2}kx}{(\epsilon^{2}+\sin^{2}kx)^{2}},\label{eq:Vna}
\end{equation}
where $\epsilon\ll1$ {[}see also Fig.\ref{fig:Vna}(a){]}. 

The fourth class considered in this work is the combined potential
\begin{equation}
V_{\textrm{comb}}(x)=V_{na}(x)+V_{\textrm{latt}}(x).\label{eq:Vcomb}
\end{equation}

All the above potentials share a common feature that the unit cell
contains one or two potential minima (as detailed in Sections \ref{subsec:classicalOpticalLattice}-\ref{subsec:classicalComb}).
Then the low lying bands $s,p,d\ldots$ are \textit{generally} simple,
single-valued bands separated by a nonzero gap. The Wannier functions
corresponding to different bands spatially overlap. The notable exception
is the double well potential in the regime when it has two very deep
potential wells per period. In this work we are interested however
in the case when $s$ and $p$ band spatially overlap.

For a ultracold gas in the lattice $p$-band, a possible loss mechanism
is when collisions push particles to different bands: one to $s$-band
and one to the $d$-band. This process may be (nearly) resonant, when
the wells of $V_{\textrm{opt.}}(x)$ are well approximated by a harmonic
potential. 

Let us consider two weakly-interacting particles in the $p$-band
with energy in the interval $\mathcal{A}=[2\min E^{1}(q),2\max E^{1}(q)]$
and collisional population of the final state with energy in ${\cal B}=[\min E^{0}(q)+\min E^{2}(q),\max E^{0}(q)+\max E^{2}(q)]$.
The process $p+p\to s+d$ is not resonant only if the distance between
intervals ${\cal A}$ and ${\cal B}$ is much larger than the interaction-induced
coupling, namely when 
\begin{eqnarray}
f\equiv\textrm{dist}({\cal A},{\cal B}) & \approx & ||2\bar{E}^{1}-\bar{E}^{0}-\bar{E}^{2}|\!-\nonumber \\
 &  & -2\left|2|J^{1}|\!+\!|J^{0}|\!+\!|J^{2}|\right|\!\gg\!U_{nnnn}^{1120}.\label{eq:fFactor}
\end{eqnarray}
The ratio $f/U_{nnnn}^{1120}$ thus quantifies the extent to which
the description of $p$-band gas given by Eq. (\ref{eq:BH}) is valid.
At the same time one desires that the parameters $U,J$ in Eq. (\ref{eq:BH})
are as large as possible in the absolute terms to make the system
stable against thermal fluctuation. Additionally the time-scales defined
by the model should be within coherence time of possible experimental
setups.

In the forthcoming Sections we show the calculation of the parameter
$f$ of the four potentials Eq.(\ref{eq:Vlatt})-(\ref{eq:Vcomb}).
We also remind the familiar relevant energy scales for hopping, interaction
and band separation for the four potentials.

\begin{figure}
\includegraphics[width=8.4cm]{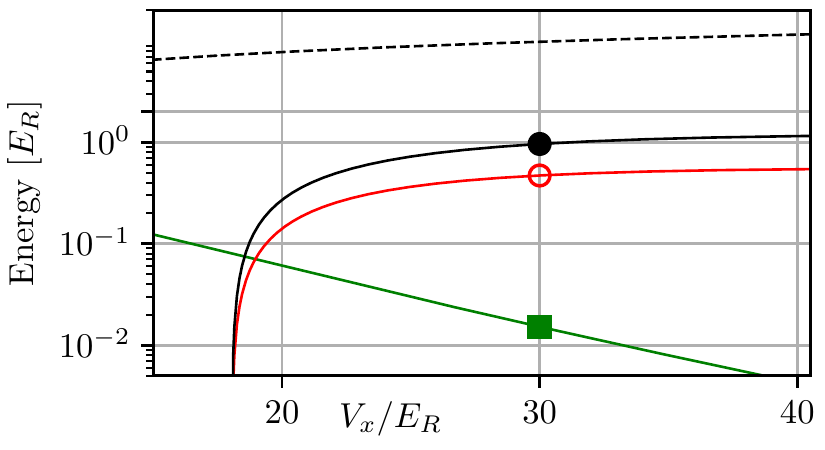}

\caption{Figure shows the value of $f$ (black curve, marked with full circle),
$\Delta_{sp}$ (black dashed line), and $U^{1111}$ (red solid line,
marked with empty circle) assuming $U^{1120}=f/10$, $J^{1}$ (green
curve, marked with a full square), for the classical optical potential
$V_{\textrm{latt}}$ as the function of its height $V_{x}$ -- see
Eq. (\ref{eq:Vlatt}).}

\label{fig:Vlatt}
\end{figure}

\subsection{Standard optical lattice potential}

\label{subsec:classicalOpticalLattice}

The low-energy sector of a Hamiltonian {[}Eq. (\ref{eq:2ndquantizH}){]}
describing a particle in the lattice potential $V_{\textrm{latt}}(x)$
{[}Eq. (\ref{eq:Vlatt}){]} for $V_{x}\gg E_{R}$ resembles a system
with a series of almost-decoupled nearly-harmonic traps. Taking into
account quartic terms in expansion of $\cos^{2}$ potential around
its minima gives an approximate formula for $n$-th band energy:
\begin{equation}
E_{n}\approx\left(n+\frac{1}{2}\right)\sqrt{4V_{x}E_{R}}-E_{qu},\quad E_{qu}=\frac{6n^{2}+6n+3}{12}E_{R}.\label{eq:Vlttcorr}
\end{equation}
The above expression is frequently used when truncated to just the
first term $\sim\sqrt{V_{x}}.$ While it is often sufficient, we note
that in $V_{x}\to\infty$ limit the quartic contribution $E_{qu}$
is non-vanishing. The $f$ defined in Eq.~(\ref{eq:fFactor}) for
very large $V_{x}\gg E_{R}$, when all the hopping rates $J^{0},J^{1},J^{2}$
are exponentially suppressed is therefore:

\begin{equation}
f\approx2E_{1}-E_{0}-E_{2}\to1E_{R}\label{eq:vlatt1er}
\end{equation}
It is important to stress that the limit value in Eq. (\ref{eq:vlatt1er})
is nonzero solely due to the quatric term in Eq.(\ref{eq:Vlttcorr}).

Specifically the $f$ is nonzero {[}see Fig. \ref{fig:Vlatt}{]} for
$V_{x}\gtrapprox18E_{R}$ At around $V_{x}\approx30E_{R}$, $f$ reaches
$1E_{R}$. The maximum value of $f$ is attained near $V_{x}\approx40E_{R}$
where $f\approx1.4E_{R}$. Increasing the lattice height $V_{x}$
and therefore $f$ comes at the price of decrease of $J^{1}.$ For
$V_{x}\approx30E_{R}\,\,(40E_{R})$ we have $J^{1}\approx0.015E_{R}\,\,(0.0042E_{R}).$

These $p$-band hopping rates are quite small. For a popular atom
species used often in experiments, the $^{87}\textrm{Rb}$, typical
parameters of lasers, the hopping frequency would be of the order
of few tens of Hertz.

\subsection{``Double-well'' potential}

\label{subsec:Double-well-potential}

By using two lasers one can create, by an AC-Stark shift a potential
of the form (\ref{eq:Vdw}). It has been used in countless theoretical
and experimental works (just a few examples being \citep{Anderlini2006,Stojanovic2008,Lee2007,Sebby-Strabley2006,Anderlini2007}).
\begin{figure}
\includegraphics[width=8.4cm]{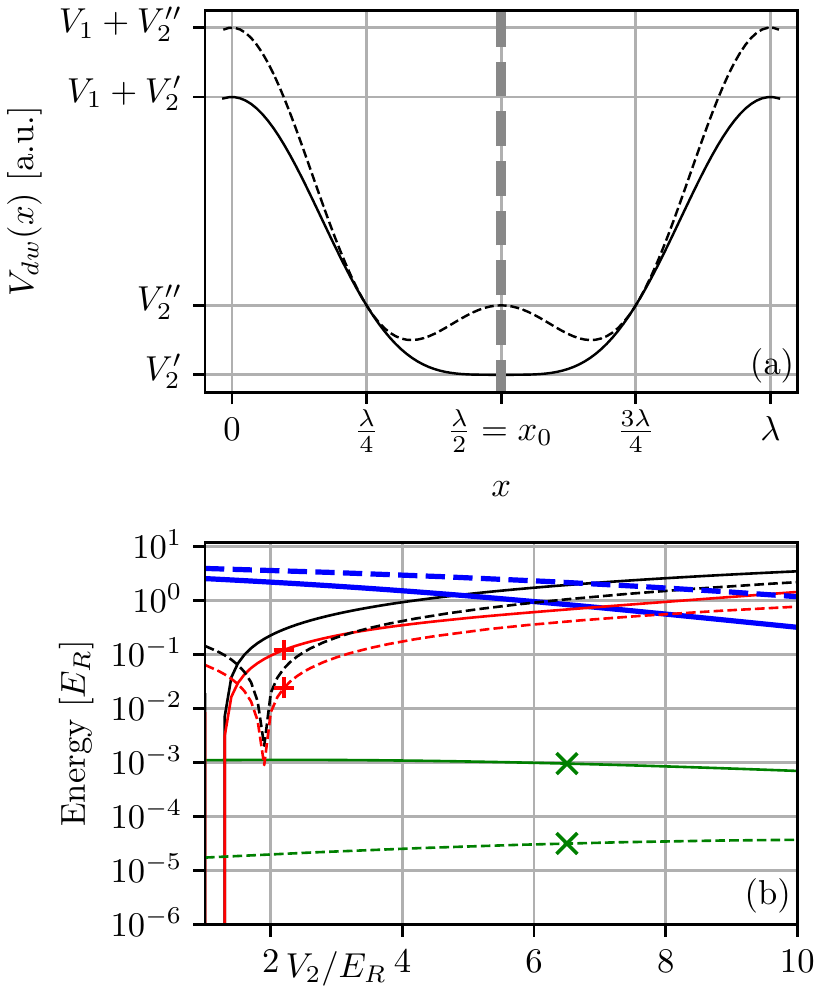}

\caption{Figure shows the properties of the potential $V_{dw}$. Panel (a)
parallels the discussion of the potential shape and symmetry in the
main text; the vertical dashed gray line shows the symmetry axis.
Solid line shows $V_{dw}$ for $V_{2}=V_{2}'=V_{1}/4$. Dashed line
shows $V_{2}=V_{2}''=V_{1}/2$ case. Panel (b) shows the value of
$f$ (black curve), $\Delta_{sp}$ (blue, thick line), and $U^{1111}$
(red with a '+'), $J^{1}$ (green with a '$\times$') assuming $U^{1120}=f/10$,
for the two-well optical potential $V_{dw}$ as the function of $V_{2}$
with $V_{1}=10E_{R}$ (solid lines) or $V_{1}=20E_{R}$ (dashed lines).
See Eq.~(\ref{eq:Vdw}).}

\label{fig:Vldoublewell}
\end{figure}

The double-well potential $V_{dw}$ has a fundamental period twice
larger than $V_{\textrm{latt}}(x)$, namely $a=\lambda$. When $\phi=0$
the potential is symmetric with respect to the center of the properly
chosen unit cell $x_{0}$ {[}see Fig. \ref{fig:Vldoublewell}(a){]}.
The Wannier functions of $s,p,d\ldots$ band are then alternately
even an odd with respect to the $x_{0}.$ For $V_{2}\leq V_{1}/4$
the potential remains unimodal within the unit cell with minimum at
$x_{0}$. At exactly $V_{1}=V_{2}/4$ at $\partial_{xx}V_{dw}(x_{\textrm{min}})=0$,
and the potential around the minimum is well-approximated by a quartic
potential {[}see Fig. \ref{fig:Vldoublewell}(a){]}. For larger $V_{2}$
there are two minima and the $s$ to $p$ band gap is significantly
reduced.
\begin{figure}
\includegraphics[width=8.4cm]{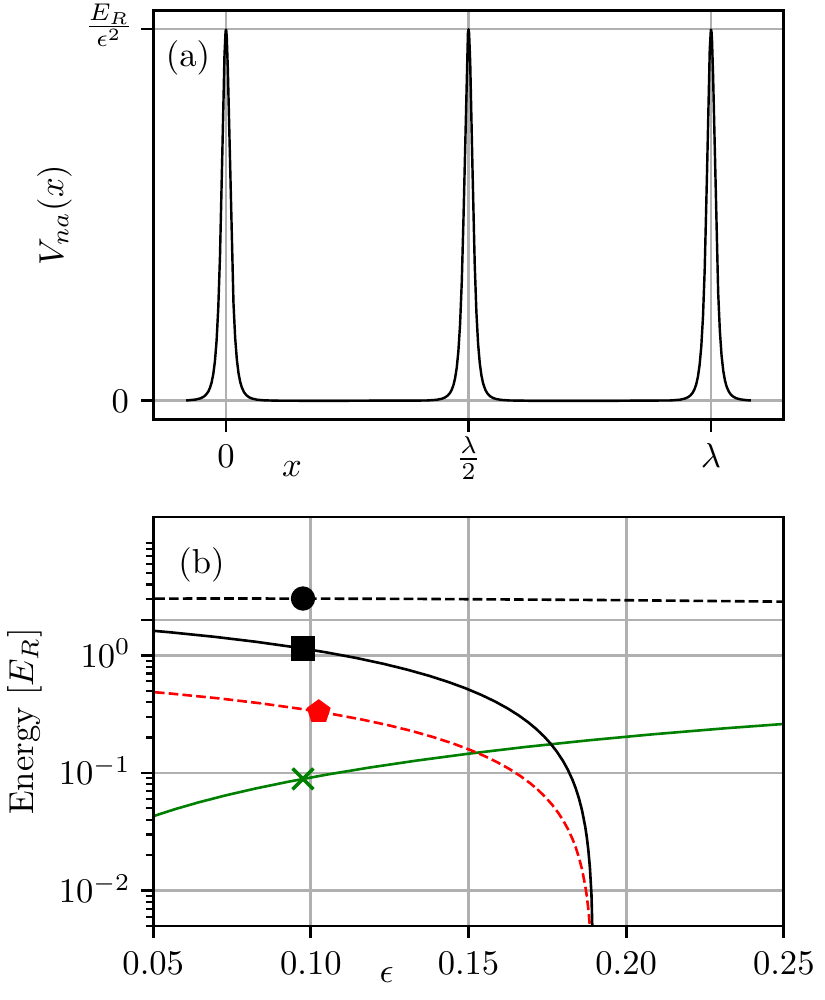}

\caption{Figure shows the properties of the potential $V_{na}$. Panel (a)
presents the shape and height of a potential $V_{na}$ parametrized
by $\epsilon$. Panel (b) shows the value of $f$ (black curve with
a square), $\Delta_{sp}$ (black dashed line with a circle), and $U^{1111}$
(red dashed line with a pentagon), $J^{1}$ (green with a cross) assuming
$U^{1120}=f/10$, for the comb optical potential $V_{na}$ as the
function dimensionless parameter $\epsilon$ -- see Eq.~(\ref{eq:Vna}).}

\label{fig:Vna}
\end{figure}

The Fig. \ref{fig:Vldoublewell}(b) shows the dependence of factor
$f$ as a function of $V_{2}$ for fixed $V_{1}=10E_{R}$ (solid lines)
and $V_{1}=20E_{R}$ (dashed lines). We see that for $V_{2}\approx V_{1}/4$
the factor $f$ is $0.35E_{R}$ for $V_{1}=10E_{R}$ and $f\approx0.65E_{R}$
for $V_{2}=20E_{R}$ the values of $f$ reached are not larger than
those achievable for the standard optical lattice ($f\leq1.4E_{R})$.
This is primarily due to twice larger period of the unit cell. As
a result the increased anharmonicity due to the dominant quartic term
is scaled down by factor 4, $E_{R}'=\frac{h^{2}}{8ma^{2}}=\frac{E_{R}}{4},a=\lambda$.
The value of $f$ can be increased by increasing both $V_{1}$ and
$V_{2}$. For fixed $V_{2}=V_{1}/4$, in the limit of very large $V_{1},V_{2}$
the factor $f$ can be arbitrarily large (in contrast to the $V_{\textrm{latt}}$).
The price to pay is the exponential decay of $J^{1}$, which for $V_{1}=10E_{R}$
is $O(10^{-3}E_{R})$ and for $V_{1}=20E_{R}$ is $O(10^{-5}\div10^{-4}E_{R}).$
Increasing just $V_{2}$, while formally leading to the increase of
$f$, actually leads to the degeneration of $p$ and $s$ band.

\subsection{Comb potential}

\label{subsec:Comb-potential}

In this section we review in detail the construction of the comb potential
(the presentation follows \citep{Lacki2016}) with a particular focus
on the parameters $\epsilon$ and $x_{0}$ which control the height
and the position of the potential peak. The details of the construction
are then referred to in discussion of $V_{\textrm{comb}}$ in the
Subsection \ref{subsec:classicalComb}.

The potential $V_{\textrm{\textrm{na}}}(x)$ as given by Eq. (\ref{eq:Vna})
transpires in the three level system \citep{Lacki2016}:

\begin{equation}
H_{\textrm{c}}=-\frac{\hbar^{2}}{2m_{a}}\partial_{x}^{2}+H_{\Lambda}(x),\label{eq:hamiltonianqg1g2e}
\end{equation}
where
\begin{equation}
H_{\Lambda}(x)=\hbar\left(\begin{array}{ccc}
0 & \Omega_{c}(x)/2 & 0\\
\Omega_{c}(x)/2 & -\Delta-i\Gamma/2 & \Omega_{p}/2\\
0 & \Omega_{p}/2 & 0
\end{array}\right)\label{eq:threechannel}
\end{equation}
has been written in the atomic level basis $|g_{1}\rangle,|e\rangle,|g_{2}\rangle.$
The Rabi frequency $\Omega_{c}(x)$ is a standing wave, 
\begin{equation}
\Omega_{c}(x)=\Omega_{c}\sin[k(x-x_{0})],\label{eq:standingwave}
\end{equation}
and $\Omega_{p}$ is $x$-independent,
\begin{eqnarray}
\epsilon & = & \frac{\Omega_{p}}{\Omega_{c}}.\label{eq:epsilon}
\end{eqnarray}
The position-dependent eigenstates of $H_{\Lambda}(x)$ include the
dark state $|f_{1}(x)\rangle=-\cos\alpha(x)|g_{1}\rangle+\sin\alpha(x)|g_{2}\rangle$
with $\alpha(x)=\arctan[\Omega_{c}(x)/\Omega_{p}]$. It is an eigenstate
to the zero energy for all $x$.

The Hamiltonian $H_{c}$ is well approximated by the following Hamiltonian
(with $V_{na}$ given by Eq. \ref{eq:Vna}):

\begin{eqnarray}
H_{na} & = & -\frac{\hbar^{2}}{2m_{a}}\partial_{x}^{2}+V_{na}(x),\label{eq:fullVna}
\end{eqnarray}
when it acts on wavefunctions of the form $\psi_{D}(x)=g(x)|f_{1}(x)\rangle$
of low energy $\langle\psi_{D}|H_{na}|\psi_{D}\rangle$ .

The potential $V_{na}$ has the form of sharp potential peaks {[}see
\ref{fig:Vna}(a){]} of height $E_{R}/\epsilon^{2}$ and width $\sim\epsilon\lambda/2\pi$,
located at $x=n\lambda/2+x_{0},n\in\mathbb{Z}.$ 

When $\epsilon\to0$ the potential peaks can be replaced by $\frac{\pi}{2\epsilon}\sum_{n}\delta(x-x_{0}-n\lambda/2).$
In that limit, mean band energies converge to values characteristic
for a box potential, $\bar{E}_{n}\approx n^{2}E_{R}.$ It may be shown
that the bandwidths of each band $\Delta E_{n}\approx4J^{n}\sim n^{2}\epsilon E_{R}$
{[}see \citep{Lacki2016}{]}. 

The potential $V_{na}$ allows to reach similar values of $f$ as
the standard optical lattice $V_{\textrm{latt}}$ ($f\approx1.4E_{R}$
for $\epsilon=0.05$, and $f\approx1E_{R}$ for $\epsilon=0.1$) {[}see
Fig. \ref{fig:Vna}(b){]}. The anharmonicity of the mean band energies
is $|2\bar{E}_{p}-\bar{E}_{s}-\bar{E}_{d}|\approx2E_{R}$ but only
for very small $\epsilon\approx0$, the hopping $J^{0},J^{1},J^{2}$
is suppressed. Experiments conducted to this date operated with values
of $\epsilon\geq0.05$ \citep{Wang2018,Tsui2019}. Further reduction
would necessitate using strong lasers creating the $\Omega_{c},\Omega_{p}$
which would couple the system beyond three levels included in (\ref{eq:threechannel}).
Use of atoms with trivial hyperfine structure would be a possible
solution to te issues raised with first experiments. It should be
also noted that the above $\Lambda$ system has not been so far experimentally
realized with the bosonic species.

The values of hopping amplitude $J^{1}$ for $\epsilon\geq0.05$ are
actually order of magnitude larger than $J^{1}$ for $V_{\textrm{latt}}$
for same value of $f$ ($J^{1}\approx0.09E_{R}$ for $\epsilon=0.1$,
and for $\epsilon=0.05$ it is $J^{1}\approx0.042E_{R}$). In contrast
to $V_{\textrm{latt}}$ it is challenging to significantly reduce
$J^{1}$ as the the potential barrier height $\sim\epsilon^{-2}$
increased with $\epsilon$ is partially compensated by its decreasing
width $\sim\epsilon$. In the subwavelength comb the $s$ to $p$
and $s$ to $d$ band separations are almost $\epsilon$-independent
and $\Delta_{sp}\approx3E_{R},\Delta_{pd}\approx5E_{R}$ .

\begin{figure}
\includegraphics[width=8.4cm]{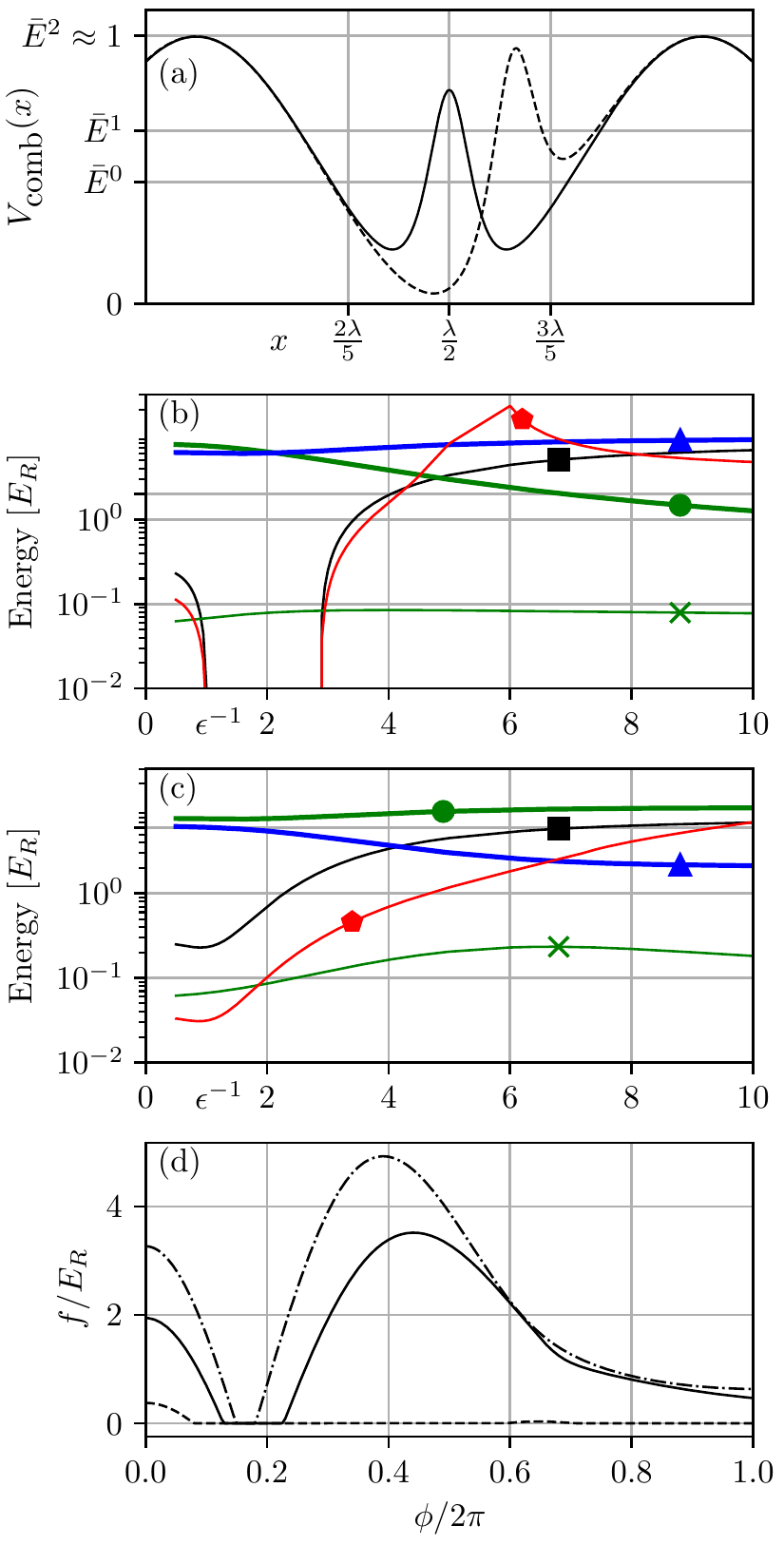}

\caption{Properties of $V_{\textrm{comb}}$. Panel $(a)$ shows the potential
shape, around the potential minimum for $V_{x}=20E_{R},\epsilon=0.25$
and $\phi=0$ (black solid line), $\phi=0.4/\pi$ (black dashed line).
The levels $\bar{E}^{0}$,$\bar{E}^{1}$,$\bar{E}^{2}$ indicate first
three band energies for $\phi=0$. Panels $(b)$ and $(c)$ show the
value of $f$ (black curve with a square), $\Delta_{sp}$ (green,
thick line with a filled circle), $\Delta_{pd}$ (blue, with a triangle
pointing up), and $U^{1111}$ (red with a pentagon), $J^{1}$ (green
with a cross) assuming $U^{1120}=f/10$, for the combined optical
potential $V_{\textrm{comb}}$ (comb potential and classical $\sin^{2}$
potential) as the function of dimensionless parameter $\epsilon$
controlling the height of the comb potential with $V_{x}=20E_{R},\phi=0$
{[}panel $(b)${]} or $V_{x}=20E_{R},\phi=0.4/2\pi$ {[}panel $(c)${]}.
See Eq.~(\ref{eq:Vdw}). Panel $(d)$ shows $f/E_{R}$ for $V_{x}=10,20,30E_{R}$
and $\epsilon^{-1}=4$ (dashed, solid, dash-dotted black line) as
a function of a shift $\phi$ of subwavelength peak w.r.t to the potential
minimum.}

\label{fig:Vcomb}
\end{figure}

\subsection{Classical potential together with the comb potential}

\label{subsec:Classical-potential-together}

\label{subsec:classicalComb}

Another possibility is opened by a combined potential that includes
both a standard lattice potential $V_{\textrm{latt}}$ and the $V_{na}$
potential (parameter $\phi$ allows for shift of $V_{na}$ peak with
respect to the minimum of $V_{\textrm{latt}}$) \citep{Budich2017}:
\begin{equation}
V_{\textrm{comb}}(x)=V_{\textrm{latt}}(x)+V_{na}(\epsilon,x-a\phi),\quad a=\lambda/2\label{eq:combinedV}
\end{equation}
The combined potential $V_{\textrm{comb}}(x)$ can be achieved when
in Eq.~(\ref{eq:threechannel}) atoms in either of states $g_{1}$
and $g_{2}$ feel an additional, standard lattice potential due to
AC-Stark shift, namely when the three-level Hamiltonian is of the
form \citep{Wang2018}.

\[
H_{\Lambda}(z)=\hbar\left(\begin{array}{ccc}
V_{1}\sin^{2}(kx) & \Omega_{c}(x-a\phi)/2 & 0\\
\Omega_{c}(x-a\phi)/2 & -\Delta-i\Gamma/2 & \Omega_{p}/2\\
0 & \Omega_{p}/2 & V_{1}\sin^{2}(kx)
\end{array}\right).
\]
The potential $V_{1}\sin^{2}(kx)$ is then simply added to $V_{na}(x-a\phi)$
resulting in the desired $V_{\textrm{comb}}(x)$.

In the pure $V_{\textrm{latt}}(x)$ potential the Wannier functions
are alternately even an odd w.r.t to the center of the unit cell.
A sharp potential in the middle of the cell, would primarily shift
energies of each band, by a value $\int_{\mathbb{R}}V_{na}(x)|W^{\alpha}(x)|^{2}dx\approx\pi|W(0)|^{2}/(2\epsilon)$.
The integral is maximal for the $s-$band Wannier functions, and decays
for subsequent even Wannier functions. It is zero for odd bands, including
the $p$-band. As the anharmonicity, for deep lattice is given by
$f\approx2\bar{E}^{1}-\bar{E}^{0}-\bar{E}^{2}>0$, adding the central
$V_{na}$ peak should first lower the $f$ first towards zero. Only
after that the $f$ can attain, large negative values. 

Non-central placement of the $V_{na}(\epsilon,x)$ offers more flexibility
in manipulating the $f\approx2\bar{E}^{1}-\bar{E}^{0}-\bar{E}^{2}$
(see also \citep{Budich2017}). For example, when the subwavelength
peaks $V_{na}$ coincide with maxima of the $p$-band Wannier function
of $V_{\textrm{latt}}(x)$ one can expect the mean $\bar{E}^{1}$
to increase strongly in contrast to $\bar{E}^{0}$ and $\bar{E}^{2}$.
As a result the band anharmonicity should be increased to large positive
values.

In Fig. \ref{fig:Vcomb}(a) shows the parameters describing the properties
of the $V_{\textrm{comb}}$ for $\phi=0$ and for $\phi=0.4/\pi$.
In the latter case $V_{na}(x-a\phi)$ nearly coincides with maximum
of the $p$-band Wannier function.

When $\phi=0$ it is evident that the values of $f$ that easily achieve
value of $f=2E_{R}$ (for $\epsilon\approx0.25)$ before, for even
smaller $\epsilon,$ the central peak given by $V_{na}$ cuts the
potential well into two almost disconnected parts, with tiny $\Delta_{sp}$.
For $\phi=0$ we focus on $\epsilon\approx0.25$ when the $\bar{E}^{1}$
is similar to the $V_{na}$ peak height, the Wannier function of the
$p$ band is not strongly affected by the potential peak and $\Delta_{sp}$
remains sizable.

For small $\epsilon^{-1}$ the potential $V_{\textrm{comb}}$ can
resemble the potential $V_{dw}.$ The crucial difference is that its
period remains $a=\lambda/2$ in contrast to period $\lambda$ of
$V_{dw}$. This allows to maintain much higher hopping rate in $V_{\textrm{comb}.}$ 

When $\phi=0.4/\pi$ the values of $f$ that are reached are similar,
with important difference: the value of $f=2\bar{E}^{1}-\bar{E}^{0}-\bar{E}^{2}$
for $\epsilon^{-1}\to0$ does not change the sign as $\epsilon^{-1}$
is increased. This means that in contrast to $\phi=0$, also for $\epsilon^{-1}\leq3$
the $f$ is nonzero. This is important for practical applications
as the spontaneous emission losses quickly grow with $\epsilon^{-1}$
(see \citep{Lacki2016,Wang2018}). 

For all considered $\phi$ the hopping rate $J^{1}\approx0.1\div0.15E_{R}$.
It is half of the order of magnitude larger than for $V_{\textrm{latt}}$of
the same potential height. This is simply due to the fact that putting
extra potential $V_{na}$ in the potential well makes it effectively
shallower. In contrast to using a standard optical lattice $V_{\textrm{latt}}$
working with the value of $f\approx1-2E_{R}$ requires $V_{x}=15\div20E_{R}$
while for $V_{\textrm{latt}}$ the $f\approx1-1.4E_{R}$ is reached
for deep lattices of $V_{x}=30\div40E_{R}$ with $J^{1}$ hopping
rate significantly reduced. At such value of $\epsilon$ the combined
potential retains workable features of the $V_{\textrm{latt}}$ such
as sizable $\Delta_{sp}\approx3.8E_{R}$.

\section{Collisional stability of p-band condensate}

\label{sec:Stability-of-p-band}

\begin{figure}[t]
\includegraphics[width=8.4cm]{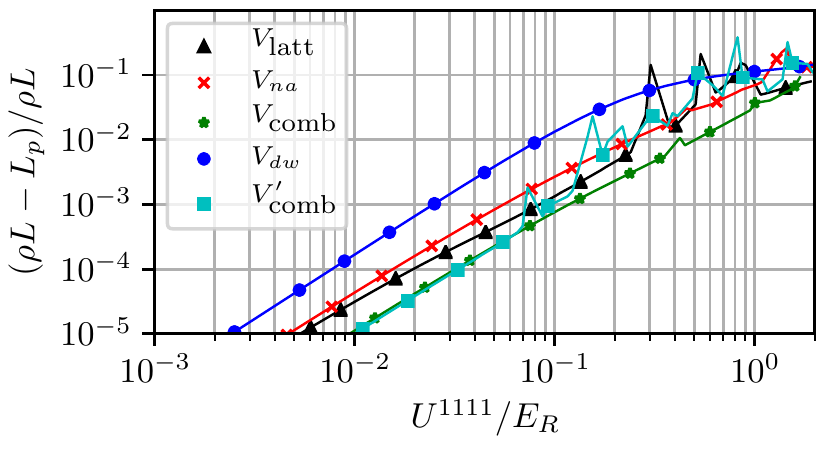}

\caption{Mean losses from the $p$-band population of an interacting system
for the four considered potentials, in a system of length $L=6$ populated
with $\rho L=7`$ particles. The exact parameters are given by Eqs.
(\ref{eq:paramlatt})-(\ref{eq:paramdw}).}

\label{fig:Np}
\end{figure}
In this section we numerically study the stability of the $p$-band
gas in the potentials from the previous Section~\ref{sec:potentials}.
To this end we consider a small, strongly interacting system consisting
of $L=6$ sites described by Hamiltonian $H_{\textrm{MBH}}$ in Eq.
(\ref{eq:MBH}) under periodic boundary conditions, restricted to
lowest three bands, $\alpha_{\textrm{max}}=3.$
\begin{enumerate}
\item Potential $V_{\textrm{latt}}$ of height $V_{x}=30E_{R}$ with $f\approx1E_{R}$
and: 
\begin{equation}
(J^{0},J^{1},u,u')\approx(-0.00046E_{R},0.015E_{R},0.698,4.89),\label{eq:paramlatt}
\end{equation}
\item Potential $V_{na}$ for $\epsilon=0.1$ with $f\approx1.11E_{R}$
and: 
\begin{equation}
(J^{0},J^{1},u,u')\approx(-0.022E_{R},0.091E_{R},0.985,3.05),\label{eq:paramna}
\end{equation}
\item Potential $V_{\textrm{comb}}$ of $V_{2}=20E_{R}$, $\epsilon=0.25,\phi=0$
with $f\approx1.94E_{R}$ and: 
\begin{equation}
(J^{0},J^{1},u,u')\approx(-0.024E_{R},0.084E_{R},0.933,9.02),\label{eq:paramcomb}
\end{equation}
\item Potential $V_{\textrm{comb}}$ of $V_{2}=20E_{R}$, $\epsilon=0.25,\phi=0.4/2\pi$
with $f\approx3.38E_{R}$ and: 
\begin{equation}
(J^{0},J^{1},u,u')\approx(-0.00074E_{R},0.11E_{R},0.445,5.88),\label{eq:paramcomb-1}
\end{equation}
\item Potential $V_{dw}$ of $V_{1}=20E_{R},V_{2}=4E_{R}$ with $f\approx0.65E_{R}$
and: 
\begin{equation}
(J^{0}\!,J^{1}\!,u,u')\approx(-1.16\!\times\!10^{-6}E_{R},2.5\!\times\!10^{-5}E_{R},0.802,4.00).\label{eq:paramdw}
\end{equation}
\end{enumerate}
Initially the quantum state of the gas of $N_{p}=\rho L$ particles
is an eigenstate $\psi_{0}$ of the Hamiltonian $H_{\textrm{MBH}}$,
Eq. (\ref{eq:MBH}) with no interactions, $g=0$. The chosen initial
state is the least energy eigenstate where all the particles populate
the $p$-band -- it is a product of Bloch functions with quasimomentum
$q=q_{*}$ minimizing $E^{1}(q)$.

The subsequent evolution $\psi(t)=\exp(-iH_{\textrm{MBH}}t/\hbar)\psi_{0}$
is governed by full $H_{\textrm{MBH}}$ truncated to lowest three
bands, $\alpha<\alpha_{\textrm{max}}=3$. The population of the $p$-band
is given by
\[
N_{p}(t)=\langle\psi(t)|\hat{n}_{p}|\psi(t)\rangle
\]
Initially, $N_{p}(t=0)=\rho L$. For $t>0$ the population $N_{p}(t)$
drops due to interaction-driven coupling to other bands. The temporal
dependence shows some long-term oscillatory behavior that is attributed
to the finiteness of the system and that of the Hilbert space. To
mitigate these effects, we consider a long-time average $L_{p}=\overline{N_{p}(t)}.$

Let us discuss the dependence of $N_{p}(t)$ and $L_{p}$ on the interaction
strength $g$, that can be altered by means of a Feshbach resonance
\citep{Chin2010}.

The value of $\xi=J^{1}/U^{1111}$ parameter together with $\rho$
indicate the position in the phase diagram of the BH model. The absolute
value of $J^{1}$ {[}Eqs. (\ref{eq:paramlatt})-(\ref{eq:paramdw}){]}
is however smaller by $V_{\textrm{latt.}}$ of $V_{dw}$ by order
of magnitude or more than $V_{na}$ or $V_{\textrm{comb.}}$. To ensure
that all four cases correspond to the same quantum phase -- superfluid
state, we fix the system density to $\rho=7/6>1$. As a result all
the four systems have essentially the same probability for double
occupation of the lattice site.

Under such assumptions the relative loss defined as $\delta=(\rho L-L_{p})/\rho L$
measures the depletion of the $p$-band. As revealed by a numerical
simulation, for as long as $U^{1111}\ll f$ the losses $\delta$ for
potentials $V_{\textrm{latt}},V_{na},V_{\textrm{comb}},V_{dw}$ scale
as $\delta=A(U^{1111}/E_{R})^{\alpha}.$ Fitting the numerical data
show in Fig. \ref{fig:Np} we find that for $V_{\textrm{latt}}$ we
have $A=0.13900\pm0.00028,\alpha=1.83090\pm0.00039$; for $V_{na}$
we have $A=0.28983\pm0.00042,\alpha=1.92133\pm0.00028$; for $V_{\textrm{comb}}$
we have $A=0.093\pm9.0\cdot10^{-5},\alpha=1.9610\pm0.0050$; for $V_{dw}$
we have $A=0.94703\pm0.00054,\alpha=1.97562\pm0.00011$. The fittings,
as evident from Fig. \ref{fig:Np} differ mainly by the prefactor,
with losses for the $V_{\textrm{comb}}$ being smaller with respect
to the $V_{\textrm{latt}}$ by approximately $30\textrm{\%}$ for
same value of $U^{1111}$. Together with order of magnitude larger
$J^{1}$ potential $V_{\textrm{comb}}$ provides a compelling way
to realize a weakly interacting $p$-band superfluid.

When the offset of the subwavelength peak is introduced, for $\phi=0.4/2\pi$
which maximizes the $f$-factor. The loss rate scaling is given by
$A=0.0863\pm0.0002,\alpha=1.96956\pm0.00043$. This offers only marginal
improvement over $\phi=0$ case. As can be seen in Eqs. (\ref{eq:paramcomb})
and (\ref{eq:paramcomb-1}), while the factor $f$ is increased from
$1.94E_{R}$ to $3.38E_{R}$ the ratio of $U^{1111}/U^{1120}$ is
decreased from 9.02 down to 5.88. This means that for same $U^{1111}$
the value of $U^{1120}$ is actually larger in the $\phi\neq0$ case.
This consumes all benefits from increasing the $f$ as $f/U^{1102}$
is essentially the same in both cases.

\section{Resonant coupling of $S-$band to $P-$band}

\subsection{Synthetic ladder}

\begin{figure}
\includegraphics[width=7cm]{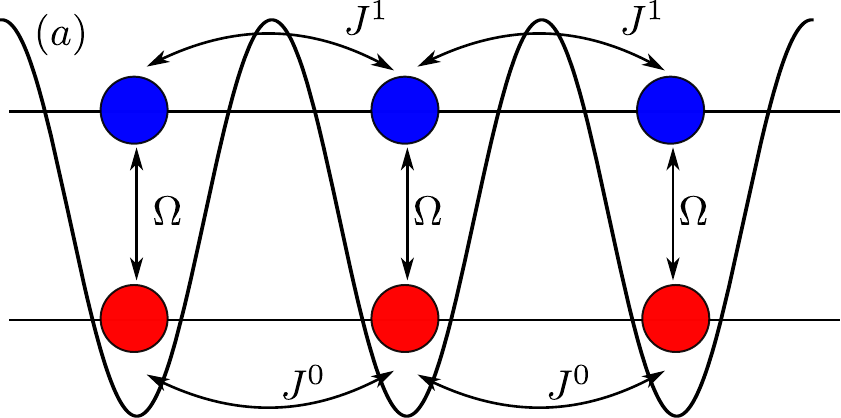}

\vspace{0.5cm}

\includegraphics[width=8.4cm]{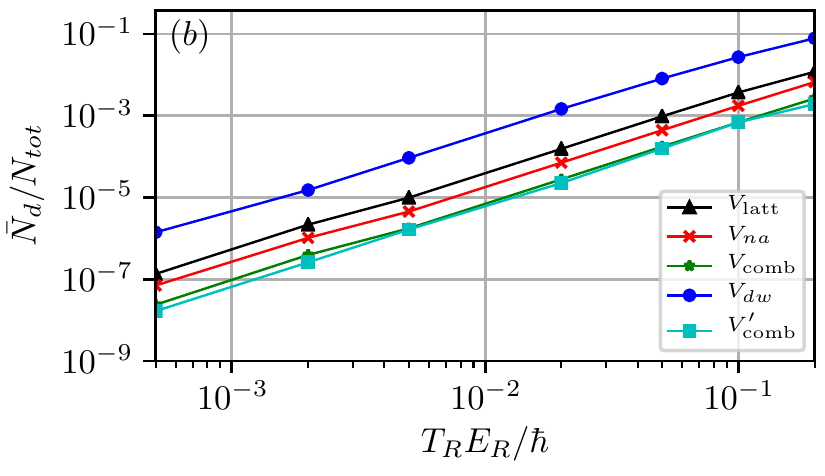}

\caption{Mean losses from $s-p$ band, noninteracting ladder system measured
as $\bar{N}_{d}$ population as a function of $T_{R}$ -- Rabi frequency
oscillation between $s$ and $p$ band (see text). Black triangles,
red crosses, blue circles, green stars show losses for $V_{\textrm{latt}},V_{na},V_{dw},V_{\textrm{comb}}$
lattices, for ($V_{x}=30E_{R},$ $\epsilon=0.1,$ $V_{1}=20E_{R},\phi=0$
$\textrm{ and }V_{2}=4E_{R},$ $V_{1}=20E_{R}$ and $\epsilon=0.25$
respectively). The cyan $V_{\textrm{comb}}'$ data series for configuration
of $V_{\textrm{comb}}$ but with $\phi=0.4/2\pi$.}

\label{fig:ladder}
\end{figure}
Coupling of the $s$ and $p$ bands can be achieved by a periodic
modulation of the system with the frequency $\omega=\bar{E}^{1}-\bar{E}^{0}$
(see \citep{Gemelke2005,Sowinski2012,acki2013,Cabrera-Gutierrez2019}).
The other bands are off-resonant if $\omega=\bar{E}^{1}-\bar{E}^{2}\neq0$
and in first approximation they can be neglected. We will study the
degree to which neglecting other bands is possible for different potentials
considered in this work.

The simplest coupling the bands $s$ and $p$ is by periodic modulation
of the position of the lattice:

\[
V(x,t)=V(x-A\sin\omega t).
\]
The potential oscillation implies a co-movement of the instantaneous
Wannier functions:
\begin{eqnarray*}
{\cal W}_{n}^{\alpha}(x,t) & = & {\cal W}_{n}^{\alpha}(x-A\sin\omega t)
\end{eqnarray*}
For fixed $t$ the Hamiltonian (\ref{eq:2ndquantizH}), in the basis
set by ${\cal W}_{i}^{\alpha}(x,t)$, will take the form of $H_{\textrm{MBH}}$
with time-independent coefficients. Nevertheless, correctly derived
time-evolution equation of motion, has to take into the account the
time-dependence of the basis. The resulting Time dependent Schr\"odinger
equation is of the form \citep{acki2013,pichler2013heating}:

\begin{eqnarray}
\partial_{t}\psi & = & H_{\textrm{MBH}}\psi-\sum_{i}\sum_{\alpha,\beta}[{\cal T}_{ii}^{\alpha\beta}A\omega\cos\omega t](a_{i}^{\alpha})^{\dagger}a_{i}^{\beta}\psi.\label{eq:tdsembh}
\end{eqnarray}
The extra terms proportional to $T_{nm}^{\alpha\beta}$ do not couple
within the same band as $\forall m,n:{\cal T}_{nm}^{\alpha\alpha}=0$.
If the potential is symmetric with respect to the middle of the unit
cell, the coefficients ${\cal T}_{ii}^{\alpha\beta}=0,\alpha+\beta\equiv0\,(\!\!\!\!\mod2),$
and they couple just bands of opposite parity, for example $s$ and
$p$ band, but also e.g. $p$ to the $d$ band. 

A 1D optical system, where population of bands other than $s$ and
$p$ is precluded can be seen as a two leg ladder system {[}see Fig.
\ref{fig:ladder}(a){]}. This shares many features with the synthetic
dimension construction where in lieu of bands, different hyperfine
states are used \citep{Celi2014}. In particular, two sites at the
ends of same ladder step, are physically in the same place, possibly
allowing for strong interaction. The hoppings along the each of the
lattice legs is governed by $J^{0}$ and $J^{1}$ respectively, possibly
allowing to implement complex ladder systems such as \citep{Li2011}.

With no interaction, such an effective single particle system decomposes
into fixed quasimomentum sectors:
\begin{eqnarray}
i\hbar\partial_{t}\left(\begin{array}{c}
\psi_{k}^{0}\\
\psi_{k}^{1}
\end{array}\right) & \!\!=H_{\textrm{ladd}} & \left(\begin{array}{c}
\psi_{k}^{0}\\
\psi_{k}^{1}
\end{array}\right),\label{eq:ladder1p}
\end{eqnarray}
where:

\begin{equation}
H_{\textrm{ladd}}=\left(\begin{array}{cc}
E^{0}(k) & iT^{01}A\omega\cos(\omega t)\\
-iT^{01}A\omega\cos(\omega t) & E^{1}(k)
\end{array}\right).\label{eq:Hladd}
\end{equation}
By transforming to a rotating frame $(\psi_{k}^{0}e^{i\delta_{0}t},\psi_{k}^{1}e^{i\delta_{1}t})$
and using the RWA, the equation of motion becomes time-independent,
when $\delta_{2}-\delta_{1}=\omega$:

\begin{eqnarray}
i\hbar\partial_{t}\left(\begin{array}{c}
\psi_{k}^{0}\\
\psi_{k}^{1}
\end{array}\right) & \!\!=\!\! & H_{\textrm{ladd},\textrm{RWA}}\!\!\left(\begin{array}{c}
\psi_{k}^{0}\\
\psi_{k}^{1}
\end{array}\right),\,\,\,\Omega=T^{01}A\omega,
\end{eqnarray}
where

\begin{equation}
H_{\textrm{ladd},\textrm{RWA}}=\left(\begin{array}{cc}
E^{0}(k)-\hbar\delta_{0} & \Omega\\
\Omega & E^{1}(k)-\hbar\delta_{1}
\end{array}\right).
\end{equation}
When $\delta_{i}\approx E^{i}(k)$, the Rabi oscillations frequency
is $\Omega=T^{01}A\omega$. For $\Omega$ dominating the bandwidths
$\Delta E^{i}$ this behavior is $k$-independent and Eq. (\ref{eq:Hladd})
can be written as:

\begin{eqnarray}
H_{L} & = & -J^{0}\sum_{n=1}^{L}(a_{n}^{0})^{\dagger}a_{n+1}^{0}-J^{1}\sum_{n=1}^{L}(a_{n}^{1})^{\dagger}a_{n+1}^{1}+H.c.\label{eq:hamLadder}\\
 &  & +i{\cal T}_{n}^{01}A\omega(a_{n}^{0})^{\dagger}a_{n}^{1}\cos\omega t+H.c.+H_{\textrm{int.}},
\end{eqnarray}
where
\begin{equation}
H_{L,\textrm{int}}=\sum_{n}\frac{U_{nnnn}^{\alpha\beta\gamma\delta}}{2}\sum_{\alpha,\beta,\gamma,\delta=0,1}(a_{i}^{\alpha})^{\dagger}(a_{i}^{\beta})^{\dagger}a_{i}^{\gamma}a_{i}^{\delta}.
\end{equation}

When coupling to other bands beyond $s$ and $p$ cannot be neglected,
system fails to be modeled by a two-leg ladder. This deviation from
the ladder system is again measured by mean occupation of $d$ and
higher bands, $\delta=\bar{N}_{\geq d}/N_{\textrm{tot}}$. Depletion
is possible also by means of the interaction.
\begin{figure}
\includegraphics[width=8.4cm]{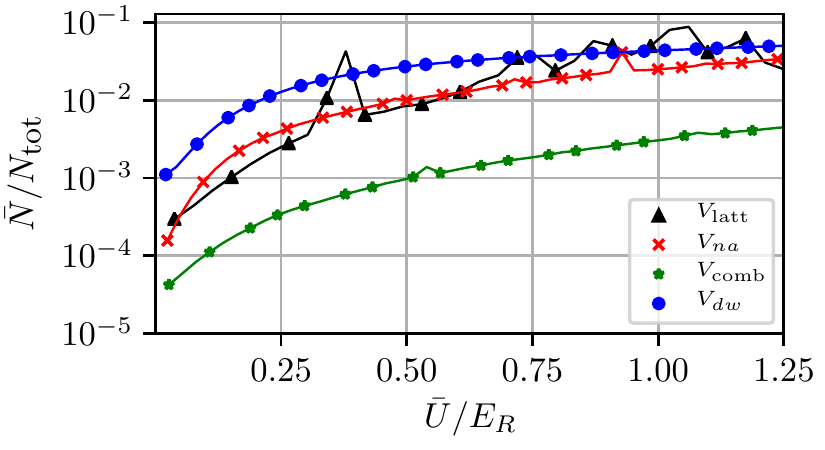} 

\caption{Depletion of the $s,p$ band by as a function of $\bar{U}=(U^{0000}+U^{1111})/2$
for such an oscillation amplitude that corresponds to $T_{R}E_{R}/\hbar=0.02$
in Fig. \ref{fig:ladder} At $\bar{U}=0$ the depletion is finite
and solely due to off-resonant coupling of $s,p$ band to higher $d$
band. Increase of the interaction results in further depletion of
$s,p$ band. The legend identifies the four potential discussed in
the main text. The $V_{\textrm{comb}}$ potential is characterized
by losses smaller for large interaction approximately by an order
of magnitude, despite much larger hopping values. }

\label{fig:ladderInt}
\end{figure}

\subsection{Realization in different potentials}

In this section we compare the implementations of Eq. (\ref{eq:hamLadder})
where the $s$ and $p$ bands are taken from the four discussed potentials.
Specifically, the ``perfect ladder'' system is Eq. (\ref{eq:tdsembh})
truncated to the $s,p$ bands. This leads to the desired Eq. (\ref{eq:hamLadder}).
The closedness of such a system is compared to the model that also
includes the $d$ band, which is the most relevant band to consider
in the study of couplings to other bands. Out of all bands beyond
$s,p$ the $\omega=\bar{E}^{1}-\bar{E}^{0}$, for the class of potentials
we consider, is closest to $\bar{E}^{2}-\bar{E}^{1}.$

To meaningfully compare the four potentials we pick the same realizations
in the Section \ref{sec:Stability-of-p-band}. In addition to parameter
values given in Eqs. (\ref{eq:paramlatt})-(\ref{eq:paramdw}), we
have that for $V_{\textrm{latt}},$$V_{na},$$V_{\textrm{comb}},$$V_{dw}$
the ${\cal T}^{01}\approx1.57E_{R},0.85E_{R},0.77E_{R},0.86E_{R}$
respectively. For the $V_{\textrm{comb}}$ potential for $\phi=0.4/2\pi$
we have $T^{01}\approx1.36E_{R}.$ It should be noted that in the
latter case, due to lack of symmetry of the unit cell, also $T^{02}\approx0.49E_{R}.$ 

We opt to compare the ladder depletion in the four potentials when
all in all four cases the Rabi oscillation period $T_{R}$ takes the
same value. Ordinarily $T_{R}=2\pi/\Omega$ with $\Omega=|T^{01}A\omega|$.
When coupling to the $d$ band is included and becomes significant,
the system does not undergo pure Rabi oscillation. More complicated
oscillation pattern emerges. We define $T_{R}$ for each amplitude
of modulation by fitting the $A+B\sin(t/T_{R})$ to the temporal dependence
of $N_{s}(t)$.

First we consider a non-interacting ladder system of total length
$L=40$. We initialize the evolution with a quasi-momentum $k=0$
state $\psi_{0}$ in the $s$-band, By fixing initial $k$ allows
consider $\Omega$ smaller than the bandwidth and still observe model
Rabi oscillations, by setting $\omega=E^{1}(k=0)-E^{0}(k=0)$. As
pictured in Fig. \ref{fig:ladder}, the losses measured by $\bar{N}_{d}$,
in the limit of small $T_{R}$ scale as $\bar{N}_{d}=A(T_{R})^{\alpha}$.
For the four potentials we find that : $\bar{N}_{d,\textrm{latt}}=(0.278\pm0.026)T_{R}^{1.909\pm0.020}$,$\bar{N}_{d,na}=(0.123\pm0.013)T_{R}^{1.899\pm0.022}$,$\bar{N}_{d,\textrm{comb}}=(0.0501\pm0.0053)T_{R}^{1.909\pm0.022}$,$\bar{N}_{d,dw}=(2.94\pm0.10)T_{R}^{1.9564\pm0.0071}$.
For the alternative implementation of $V_{\textrm{comb}}$ with $\phi=0.4/2\pi$
we get $\bar{N}_{d,\textrm{comb}}'=(0.055\pm0.0031)T_{R}^{1.974\pm0.011}$.
This scaling applies when $\bar{N}_{d}\ll1$, In all cases we find
that $\bar{N}_{d,\textrm{comb}},$ are approximately an order of magnitude
smaller than $\bar{N}_{d,\textrm{latt}},$ with $\bar{N}_{d,na},$
being in between. The losses for the double-well are order of magnitude
larger than for the remaining systems. This is because of the lowest
value of $f\approx0.65E_{R}$ out of the four samples. The potential
$V_{\textrm{comb}}$ with $\phi=0.4/2\pi$ again offers same stablity
as the $\phi=0$. Nevertheless it features a strongly suppressed $J^{0}$
hopping along one of the lattice ``legs'' {[}as implied by Eq. (\ref{eq:paramcomb})
and Eq. (\ref{eq:paramcomb-1}){]}.

If the modulated system is also interacting an interesting interplay
of shaking and interaction may occur. To simulate such a case we consider
the same four potentials, modulated with a modulation amplitude $A$
such that the noninteracting Rabi oscillation period $T_{R}=0.02\hbar/E_{R}$.
The length of the lattice is $L=6$ with total $\rho L=7$ particles. 

The Fig. \ref{fig:ladderInt} shows the dependence of the $\bar{N}_{d}$
on the mean interaction strength $\bar{U}=(U^{0000}+U^{1111})/2$
(see also (\ref{eq:paramlatt})-(\ref{eq:paramdw})). The interaction
strength would be tuned with use of the Feshbach resonance, thus scaling
both $U^{1111},U^{0000}$ by a common factor. For $\bar{U}=0$ the
results from Fig. \ref{fig:ladderInt} are reproduced. For increasing
$\bar{U}>0$ for each of the four potentials the interaction causes
the additional losses on top of the modulation losses. It is evident
that the losses for the $V_{\textrm{comb}}$ potential grow slowest
with the $\bar{U}$ {[}here again losses for $\phi=0$ are essentially
the same as for $\phi=0.4/2\pi${]}. This is due to the combination
of factors: first it allows to reach high anharmonic band placement
(large value of $f\approx2E_{R})$, second as evident from Fig. \ref{fig:Vcomb}
the value of $|U^{1102}/\bar{U}|$ is relatively small compared to
other potentials 0.17,0.32,0.11,0.23 for $V_{\textrm{latt}},V_{na},V_{\textrm{comb}},V_{dw}$
(for $V_{\textrm{comb}}$ for $\phi=0.4/2\pi$ we have $|U^{1102}/\bar{U}|\approx0.105$
). This means that if $\bar{U}$ is fixed to a given value, then the
terms responsible for interaction-driven losses come with a smaller
prefactor than in the other potentials. This advantage together with
largest value of $f$ (which affects also modulation) allows to observe
order of magnitude smaller depletion of the $s,p$ band system. This
also explains the difference in stability for $\bar{U}\neq0$.

\section{conclusions and outlooks}

The subwavelegth comb potential makes it possible to implement potentials
that break the constraints implied by the diffraction limit put on
constructions based on the AC-Stark effects. We have constructed lattice
potentials with anharmonic potential wells, that can be realistically
implemented in laboratory. The anharmonicity enhances the collisional
stability of the $p$-band gas. Moreover, the $s$ and $p$ bands
of the lattice can be resonantly coupled by modulation, and the resulting
couplings from~$s,p$ bands to other bands can be sufficiently off-resonant
to neglect them for even stronger coupling strength than the standard
lattice. The latter was presented by implementing a synthetic dimension
$s-p$ band ladder.

Another important feature of the constructed potentials is the ability
to preserve large value of the hopping rate. They offer controllable
work with the $p$ band in the regime of large hopping amplitude.

The remaining open question is the applicability of the construction
for the interacting bosonic systems. A fundamental problem is description
of interaction-driven depletion of dark states, and a choice of a
proper atom species that would allow the $\Lambda$ system construction
in for an collisionally interacting ultracold atom gas.
\begin{acknowledgments}
This work has been supported by National Science Centre project 2016/23/D/ST2/00721
and in part by PL-Grid Infrastructure (prometheus cluster).
\end{acknowledgments}

\bibliographystyle{apsrev4-1}
\input{notex_v2.bbl}

\end{document}

%% file: notex_v2.bbl
%